\documentclass[conference]{IEEEtran}

\IEEEoverridecommandlockouts

\usepackage{cite}
\usepackage{url}
\usepackage{graphicx}
\usepackage{amsmath}

\usepackage{amssymb}
\usepackage{color}
\usepackage{bm}
\usepackage{multirow}
\usepackage{algorithm}
\usepackage{algpseudocode}
\usepackage{todonotes}

\usepackage{rotating}

\usepackage{varwidth}

\usepackage{algorithm}
\usepackage{algpseudocode}
\algnewcommand\algorithmicinput{\textbf{Input:}}
\algnewcommand\INPUT{\item[\algorithmicinput]}
\algnewcommand\algorithmicoutput{\textbf{Output:}}
\algnewcommand\OUTPUT{\item[\algorithmicoutput]}

\usepackage{mathtools}

\DeclareMathAlphabet\mathbfcal{OMS}{cmsy}{b}{n}

\newcommand{\Fig}[1]{Fig.~\textup{\ref{#1}}}
\graphicspath{{figures/}}
\usepackage{subcaption}
\usepackage{textcomp}

\newtheorem{theorem}{Theorem}

\renewcommand{\epsilon}{\varepsiolon}
\renewcommand{\phi}{\varphi}

\def\BibTeX{{\rm B\kern-.05em{\sc i\kern-.025em b}\kern-.08em
    T\kern-.1667em\lower.7ex\hbox{E}\kern-.125emX}}

\begin{document}


\title{  Performance Analysis of Fronthaul Compression \\ in Massive MIMO Receiver

\thanks{The research was carried out at Skoltech and supported by the Russian Science Foundation (project no. 24-29-00189).
}}

\author{
\centering\IEEEauthorblockN{1\textsuperscript{st} Roman Bychkov}
\IEEEauthorblockA{
\textit{Skoltech}\\
Moscow, Russia \\
R.Bychkov@skoltech.ru}
\and
\IEEEauthorblockN{2\textsuperscript{nd} Andrey Dergachev}
\IEEEauthorblockA{
\textit{Skoltech}\\
Moscow, Russia \\
A.Dergachev@skoltech.ru}
\and
\IEEEauthorblockN{3\textsuperscript{rd} Alexander~Osinsky}
\IEEEauthorblockA{
\textit{Skoltech}\\
Moscow, Russia \\
A.Osinskiy@skoltech.ru} 
\and
\IEEEauthorblockN{4\textsuperscript{th} Dmitry Lakontsev}
\IEEEauthorblockA{
\textit{Skoltech}\\
Moscow, Russia \\
D.Lakontsev@skoltech.ru}
\and
\IEEEauthorblockN{5\textsuperscript{th} Andrey Ivanov}
\IEEEauthorblockA{
\textit{Skoltech}\\
Moscow, Russia \\
AN.Ivanov@skoltech.ru}
}

\maketitle
\IEEEdisplaynontitleabstractindextext

\begin{abstract}

Future generations of cellular systems presume to use an extremely high number of antennas to enable mm waves. Increasing the number of antennas requires a growth in connections between a remote radio head (RRH) and a baseband unit (BBU). Therefore, the traffic load between RRH and BBU has to grow, and the compression of interconnection between them becomes a serious problem.

In this paper, we propose a compression scheme to reduce the bitrate of the fronthaul interface that connects BBU and RRU. Then we justify compression block size and mantissa length to guarantee the required error vector magnitude (EVM). The knowledge of propagation channel sparsity and the condition number of the channel matrix helps to achieve higher compression ratios without performance loss. Simulation results with a realistic propagation channel are provided to confirm theoretical derivations. 

 \vskip 0.2cm


\end{abstract}


\section{Introduction}
\label{sec:introduction}

\IEEEPARstart {C}{ellular} network research is mostly focused on Enhanced Mobile Broadband (eMBB), which offers a fast mobile Internet \cite{shafi20175g}. The three major methods for increasing the data rate are to use a ``huge bandwidth'' band, boost the signal-to-noise ratio, and apply the technique known as ``spectrum reuse''. Massive Multiple Input Multiple Output (mMIMO) technologies rely on the principle of ``spectrum reuse'', which implies the separation of data streams in the spatial domain.

Most currently employed spectrum widths are restricted by tens of MHz due to sub-6G carrier frequency, and the radio unit's design limits the number of antennas. Thus, it is evident that in 5G communications, the strategies of a ``huge bandwidth'' and ``spectrum reuse'' have reached their limit. Therefore, for next-generation systems like 6G cellular communication networks, it makes sense to go to the higher millimeter waves, where a significantly wider bandwidth is available for data transmission. If the antenna size is reduced and the carrier frequency is raised to tens of gigahertz, a standard form factor can support several times more antennas. Thus, using an ultra-massive number of antennas (Ultra mMIMO) in the higher millimeter range makes sense, since the efficiency of the spectrum reuse will increase \cite{Rusek}.    

One of the challenges in implementing a remote radio head (RRH) with numerous antenna elements is the signal transportation to the baseband unit (BBU). Compared to the previous generation, the number of antennas has increased from $8$ in 4G to $1024$ in 5G. And the number of antennas is expected to be even higher in 6G. Moreover, the number of interconnections is expected to grow, considering the improved performance promised by Cloud Radio Access Network (C-RAN) and cell-free mMIMO \cite{Fronthaul-constrained, Efficient}.

A wireless system paradigm known as C-RAN includes real-time cloud computing, cooperative radio, centralized processing, and energy-efficient infrastructure. The entire BS is divided into BBU and RRU in C-RAN, where the connection between them is referred to as fronthaul (FH), implemented over high-bandwidth and low-latency optic fiber cables. Unfortunately, most existing FH links have insufficient capacity to transfer so many digital antenna signals from RRHs to BBUs. Thus, existing infrastructure cannot transfer information in the Ultra mMIMO system without signal compression \cite{EUSIPCO}.

 \begin{figure}[t!]
 \centering \includegraphics[width=0.95\columnwidth]{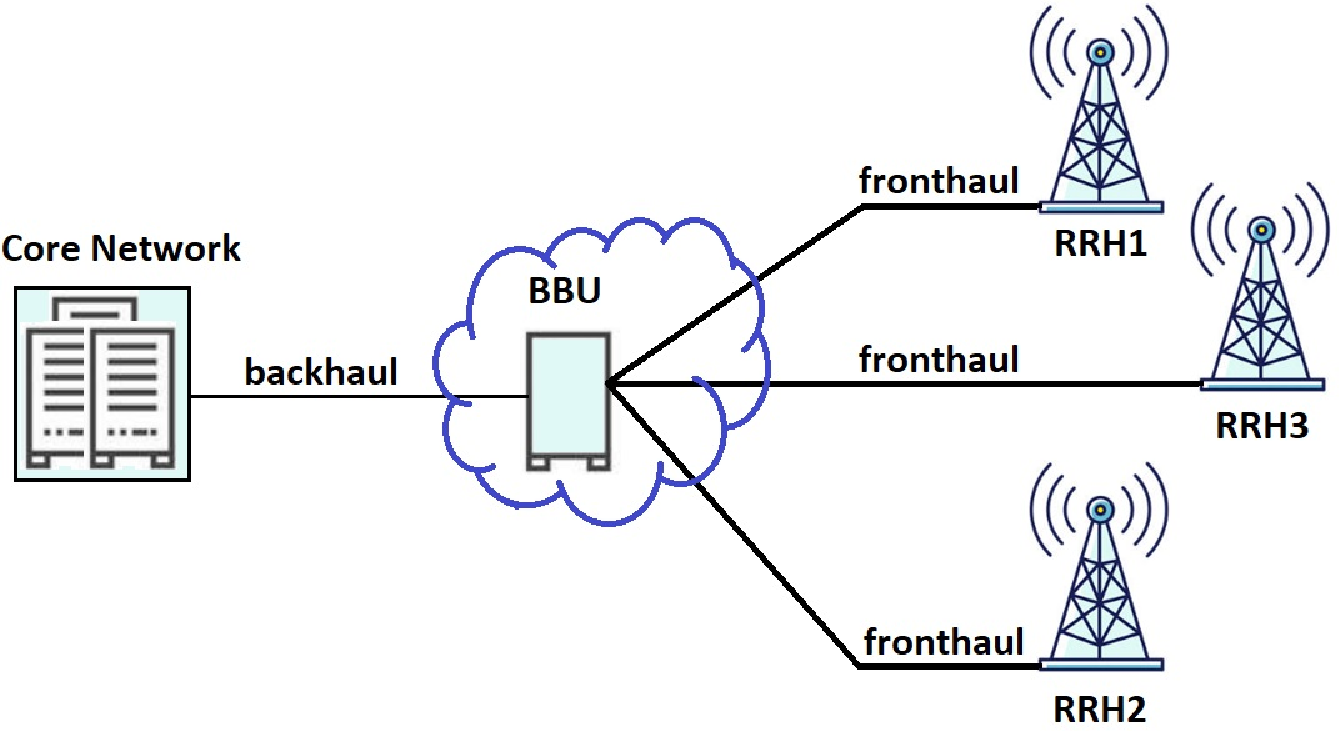}
 \caption{
 Cloud architecture: RRUs connected to BBU.
 }
 \label{network}
 \end{figure}

Massive MIMO in conjunction with C-RAN presents a significant difficulty that may limit the deployment's efficiency, despite the many benefits. In large MIMO, the FH link's capacity is a significant performance constraint for uplink reception. As more RRUs and antennas are employed to achieve enormous MIMO gains, a corresponding increase in the amount of data that must be transmitted across the FH links to the BBU. However, that for each FH stream, the higher FH load resulting from the larger antenna is not sufficiently compensated for by simply compressing the time-domain samples. Thus, decreasing the number of FH streams is necessary to increase system scalability concerning the number of antennas while also decreasing the expense of the FH network.

To overcome the overloaded FH problem, many technologies have been proposed: various functional splits \cite{IRC-Based}, distributed detection (cell-free mMIMO \cite{Björnson}), hybrid beamforming (HBF), beamspace processing, and many others. Multiple signal compression techniques have been studied: nonlinear quantization \cite{Vector}, common-exp packet, arithmetic coding, and others \cite{Multivariate}. Unfortunately, most of them lack a better theoretical justification. They don't consider propagation channel statistics and ignore channel knowledge at all. This results in bitstream overloading, worse FH compression, and therefore higher infrastructure costs.

\subsection{Our contribution}   

In this paper, we justified FH compression methods and proposed a theoretical justification for the mantissa length depending on the condition number of the propagation channel matrix. This is the major novelty compared to other existing papers, resulting in a significantly higher compression ratio.

Based on typical parameters of the 5G system, the following compression techniques were proposed for an intra-PHY functional split: signal transformation to the beamspace, common-exp packets, and non-uniform quantization. Surrogate modeling was employed to optimize parameters, considering the propagation channel sparsity. 

A realistic propagation channel generated by the 3GPP-NLOS model of Quadriga 2.0 software \cite{Quadriga} was employed to validate both compression ratio and error vector magnitude (EVM) degradation. 

\section{Compression algorithm}

Since the beginning of 5G, there is a possibility to break down the BBU to give network operators more flexibility in the FH load \cite{thesis}. 

\subsection{Functional split}
3GPP has proposed eight functional split options, at which Some BBU functions have been moved to the RRH. Split $8$ means the highest load, where FH carries time domain signals for all antennas (original C-RAN configuration). Split $1$ corresponds to the minimal FH load application (it is similar to a small cell scenario), where all services responsible for cooperative MIMO are not available, resulting in performance degradation. 

Thus, to avoid potential performance loss, we propose using intra-PHY split 7-2, where Fast Fourier Transform (FFT), cyclic prefix (CP) deletion, and digital beamforming functions are performed in the RRH. In split 7-2 the FH bit rate is dramatically decreased compared to split 7-1, while split 7–3 promises a small reduction using distributed detection at the cost of potential performance loss. Thus, split 7-2 looks reasonable. 

\subsection{Compression and decompression}\label{compress}

The compression algorithm on the RRU side can be implemented using the following steps:
\begin{enumerate}
\item Delete CP;
\item Transform signal to the frequency domain (time-to-frequency) to select only used subcarriers;
\item Transform antenna signal to the beamspace domain (antenna-to-beamspace) to reduce signal dimensionality;
\item Group signal amplitudes to common-exponent blocks (each block consists of floating point values where all mantissas have the same exponent value);  
\item Apply non-uniform quantization to mantissa inside each block to reduce mantissa bitwidth.
\end{enumerate}

\begin{figure}[t!]
 \centering
\includegraphics[width=0.95\columnwidth]{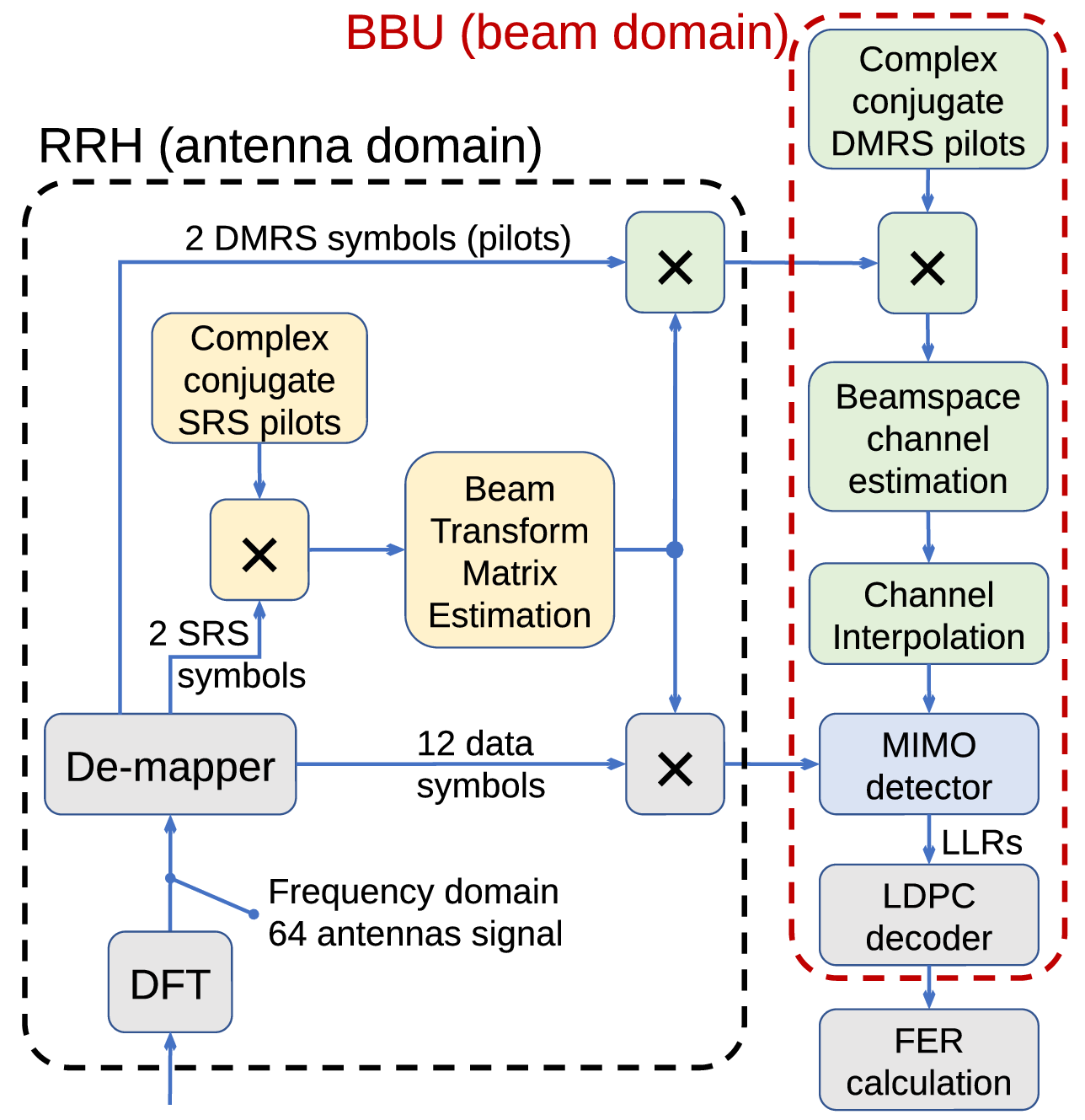}
 \caption{
 Receiver structure with the beamspace transformation.
 }
 \label{beamspace}
 \end{figure}

Step 3 (signal transformation to the beamspace domain) is a new option that appeared in mMIMO. It improves both signal compression and receiver performance \cite{Lower_bound}. In addition to dimensionality reduction, it performs spatial filtering of the user signal in a noisy environment \cite{Spatial_Denoising}. At the RRH side, the antenna signal is multiplied by the beam transform matrix selected from the most powerful spatial direction. The receiver structure utilizing processing in the beamspace domain is shown in \Fig{beamspace}.

  \begin{figure}[t!]
 \centering
\includegraphics[width=0.95\columnwidth]{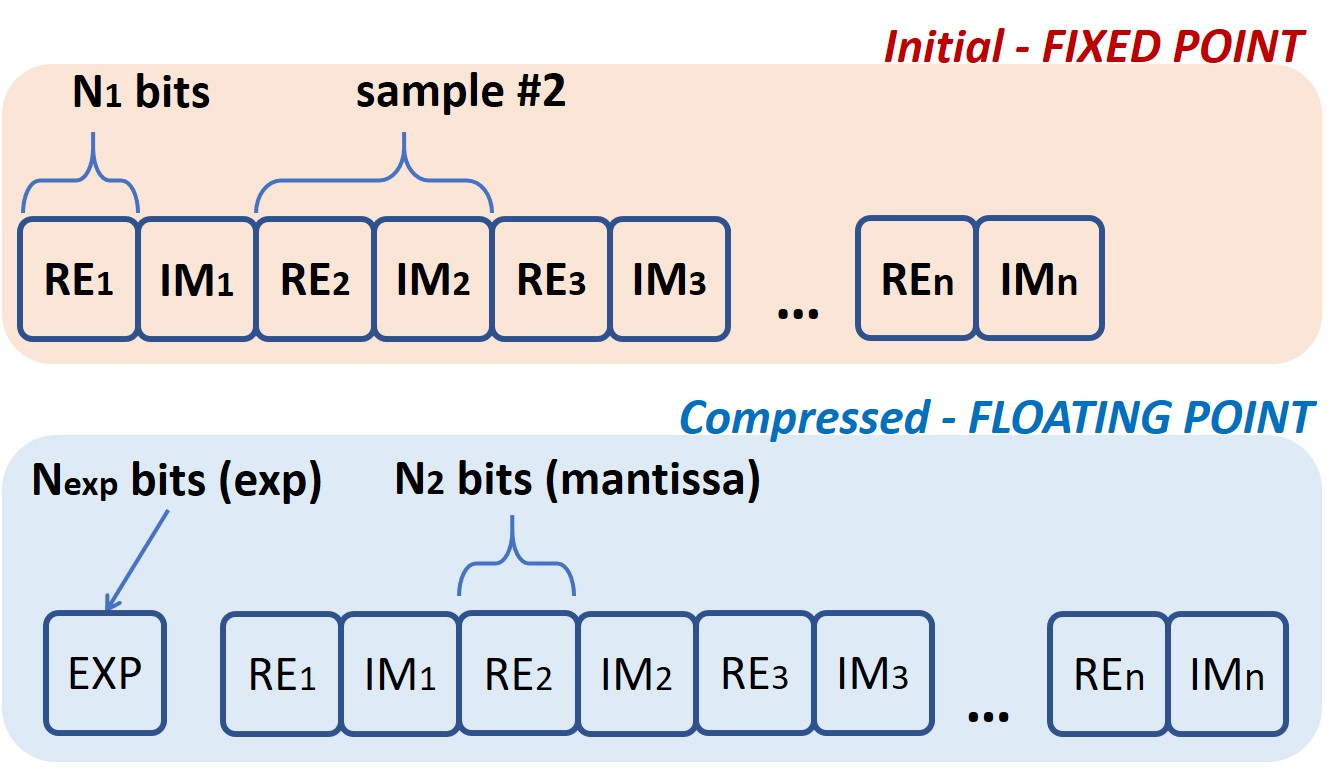}
 \caption{
 Block of floating point values with a common exponent.
 }
 \label{common_exp}
 \end{figure}

Step 4 performs the transformation of signal amplitudes from the fixed-point format to the special floating-point format having a common exponent value as shown in \Fig{common_exp}. Such blockwise division is reasonable because a scheduler shares frequency domain resources to users by elementary groups named resource blocks (RB) that consist of a set of 12, 24, or 48 subcarriers. Thus, signal power and signal-to-noise ratio (SNR) are expected to be more or less the same for subcarriers inside each RB. Therefore, using a common exponent, that performs a scaling function, looks quite reasonable. An amplitude variation inside RB is caused by modulation of subcarrier amplitudes (as they carry information) and by frequency selective fading after signal propagation through the multipath channel. 

At the BBU side, the signal is decompressed as follows:
\begin{enumerate}
\item Apply inverse non-uniform quantization to each mantissa;
\item Multiply each mantissa by the corresponding exponent value to calculate floating point values.
\end{enumerate}

Then MIMO detection and decoding are performed in the beamspace domain according to \Fig{beamspace}, without transforming back to the antenna domain. 

Thus, the major problem is to predict the impact of the FH compression noise (or roundoff error) on the resulting EVM value calculated after MIMO detection.

\subsection{MIMO detection}

Consider the frequency domain antenna signal at the MIMO receiver as follows \cite{Rusek}:
\[
Y = HX + E, 
\]
where $X \in \mathbb{C}^N$ is the transmitted signal, $Y \in \mathbb{C}^M$ is the received signal, $H \in \mathbb{C}^{M \times N}$ is the channel matrix, and $E \in \mathbb{C}^M$ is the white noise, $M$ is the number of antennas, $N$ is the number of users.

The standard linear solution for computing $X$ is given by the minimum mean squared estimate (MMSE) with white noise power $\sigma^2$ \cite{cloud}:
\begin{equation}\label{MMSE}
\begin{aligned}
  \hat X &= W Y,\\
  W &= \left( H^H H + \sigma^2 I \right)^{-1} H^H.
  \end{aligned}
\end{equation}

The MMSE detection \eqref{MMSE} requires matrices multiplication and inversion. Therefore, the roundoff error of $Y$ can significantly impact the $\hat X$ accuracy, even if the algorithm \eqref{MMSE} is implemented in high-precision arithmetic.  

\section{Error prediction - constant mantissa length}

Note, that when the noise power is low, it is close to the solution of the linear least squares (LS) problem \cite{cholscal}:
\begin{equation}\label{eq:HXY}
  \left\| HX - Y \right\|_2 \to \min,
\end{equation}
This will be even more so if we use the MMSE detection in the beamspace domain, where the SNR is even higher. In general, the solution to the LS detector can be written using the pseudoinverse $H^+$ to the channel matrix $H$:
\[
  \hat X = H^+ Y.
\]
If the noise power in $Y$ is lower than the signal power, we also have $\left\| Y \right\|_2 \sim \left\| HX \right\|_2$.

In \cite{OurCholBound} it was established that real channels show a similar roundoff to that of random (randsvd) matrices, and the corresponding error can be replaced by a random Gaussian one. It mainly relies on the following theorem.


\begin{theorem}[\cite{OurCholBound}]\label{prop:gauss}
Let $\Omega_{M \times N} \in \mathbb{C}^{M \times N}$ be a random Gaussian matrix with independent entries. Then for any matrices $A \in \mathbb{C}^{M' \times M}$ and $B \in \mathbb{C}^{N \times N'}$
\[
\begin{aligned}
  \left\| A \Omega_{M \times N} B \right\|_F \sim \frac{\left\| A \right\|_F \left\| \Omega_{M \times N} \right\|_F \left\| B \right\|_F}{\sqrt{MN}},
\end{aligned}
\]
where
\[
  \left\| A \right\|_F \sim \left\| B \right\|_F \Leftrightarrow \mathbb{E} \left\| A \right\|_F^2 \sim {\rm const} \cdot \mathbb{E} \left\| B \right\|_F^2.
\]

\end{theorem}

\subsection{MIMO detection error caused by FH compression}


Denote the roundoff error in data $Y \in \mathbb{C}^{N_{beam}}$ by $\Delta Y$. For simplicity, let us replace $\Delta Y$ by a random Gaussian vector. Then, substituting $M = N_{beam}$ and $N = 1$ into Theorem \ref{prop:gauss}, $\Omega = \Delta Y \in \mathbb{C}^{N_{beam} \times 1}$ and $B = 1 \in \mathbb{C}^{1 \times 1}$, we have
\begin{equation}\label{eq:Yprod}
  \left\| A \Delta Y \right\|_2 \sim \left\| A \right\|_F \left\| \Delta Y \right\|_2 / \sqrt{N_{beam}}
\end{equation}
for any fixed matrix $A$.

Similarly, for random data, $X \in \mathbb{C}^{N_{user}}$ we obtain
\begin{equation}\label{eq:Xprod}
  \left\| A X \right\|_2 \sim \left\| A \right\|_F \left\| X \right\|_2 / \sqrt{N_{user}}
\end{equation}
again for any fixed matrix $A$.

Let's say we save a vector $Y$ with relative error $\delta_Y = \left\| \Delta Y \right\|_2 / \left\| Y \right\|_2$. Then
\[
  \Delta \hat X = H^+ \Delta Y.
\]
This leads to, using equations \eqref{eq:Yprod} and \eqref{eq:Xprod},
\begin{equation}\label{CPRI_error}
\begin{aligned}
  \left\| \Delta \hat X \right\|_2 & \sim \left\| H^+ \Delta Y \right\|_2 \sim \frac{1}{\sqrt{M}} \left\| H^+ \right\|_F \left\| \Delta Y \right\|_2 \\
  & \sim \frac{\delta_Y}{\sqrt{N_{beam}}} \left\| H^+ \right\|_F \left\| Y \right\|_2 \\
  & \sim \frac{\delta_Y}{\sqrt{N_{beam}}} \left\| H^+ \right\|_F \left\| HX \right\|_2 \\
  & \sim \frac{\delta_Y}{\sqrt{N_{beam} N_{user}}} \left\| H^+ \right\|_F \left\| H \right\|_F \\
  & = \frac{\delta_Y}{\sqrt{N_{beam} N_{user}}} {\rm cond}_F \left( H \right).
\end{aligned}
\end{equation}
Here ${\rm cond}_F \left( H \right) = \left\| H^+ \right\|_F \left\| H \right\|_F$ is the condition number in the Frobenius norm, which is usually of the same order as the regular condition number in spectral norm due to exponential distribution of singular values of realistic channels. 

\subsection{Error growth caused by a common exponent}

As described in Step 4 of Section \ref{compress}, the same exponent value is employed for different subcarriers (elements in different vectors $Y$). The exponent value depends on the resource block (RB) index and the beam index. Thus, we should increase $\delta_Y$ accordingly, since now it's not compared to the current value of $Y$, but the maximum among $N_{12}=12$ complex values (a single RB). Since the maximum of $2N_{12}$ Gaussian random variables (we account for both real and imaginary parts) is, on average, not more than $\sqrt{2 \ln \left( 4 N_{12} \right)}$ times higher (from a simple argument using Jensen's inequality), our estimate also grows by the same factor as follows
\begin{equation}\label{mu_common_exp}
  \left\| \Delta \hat X \right\|_2 = \delta_Y  \sqrt{\frac{2 \ln \left( 4 N_{12} \right)}{N_{beam}N_{user}}}{\rm cond}_F\left( H \right).
\end{equation}

\section{Trainable mantissa length}

When replacing $\Delta Y$ with a Gaussian vector, we used the assumption that elements of $Y$ are equally distributed. Thus, equation \eqref{mu_common_exp} is always valid in the antenna domain, where the expected power on each antenna is the same. In addition, it can work in the beamspace domain in the multi-user scenario, where all or almost all beams contain user signals.

Let's analyze a case where an RB contains only a single user signal. In this scenario, $H$ is a vector, therefore, ${\rm cond}_F \left( H \right) = \left\| H \right\|_2^{-1} \left\| H \right\|_2 = 1$, and the MMSE error \eqref{CPRI_error} becomes
\begin{equation}\label{1user}
  \left\| \Delta \hat X \right\|_2 = \frac{\delta_Y}{\sqrt{N_{beam}}}.
\end{equation}

\begin{figure}[t!]
\centering
\includegraphics[width=1.00\columnwidth]{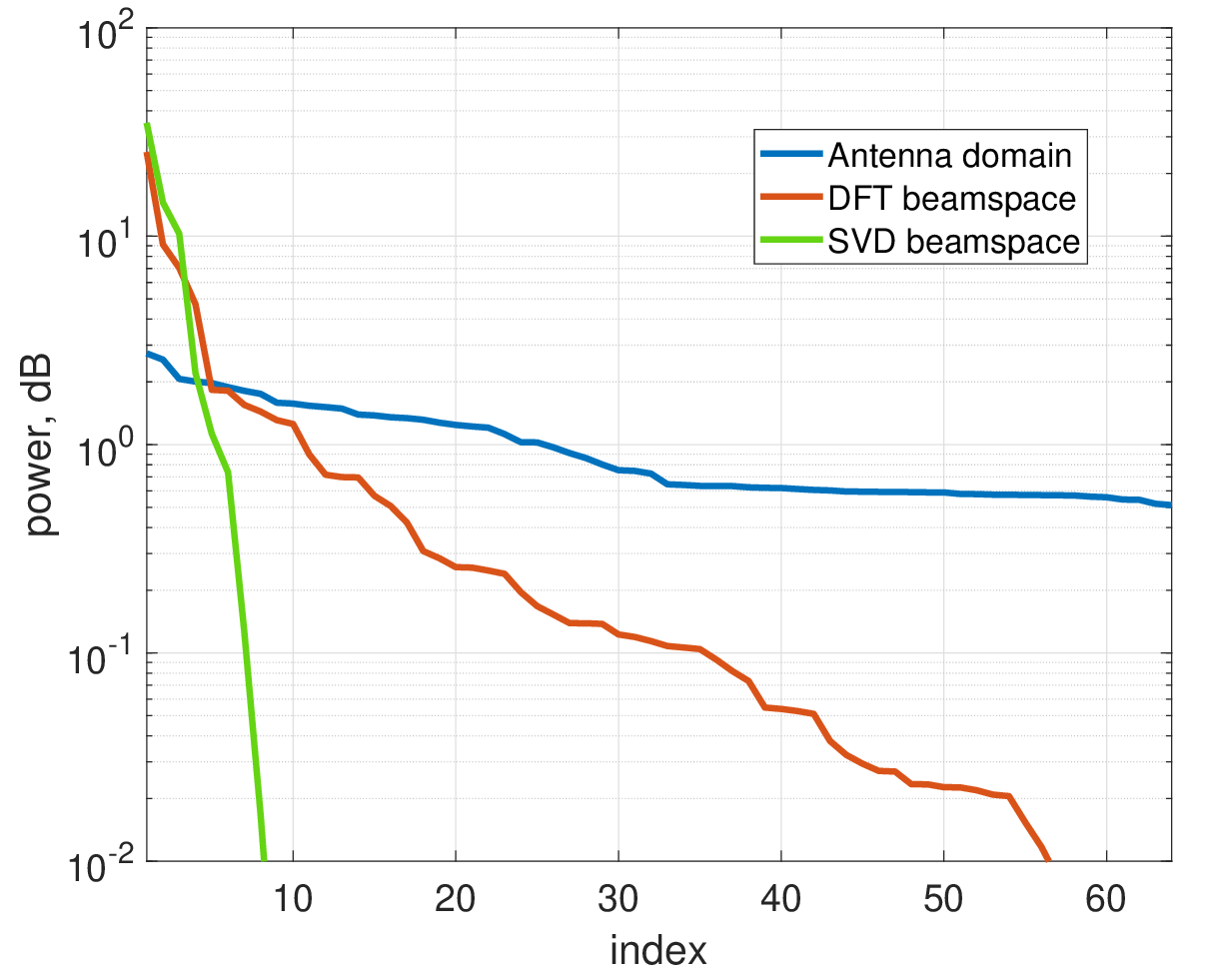}
\caption{Power distribution for a single user (antenna and beamspace domains).
}
\label{power_distribution}
\end{figure}

In practice, the average power of each element in $Y$ can vary, which can lead to a different coefficient than $1/\sqrt{N_{beam}}$. A realistic spatial distribution of the received power is presented in \Fig{power_distribution} for a single-user scenario without additive noise. It was generated by Quadriga 2.0 \cite{Quadriga} software with parameters specified in Table \ref{tab:params}. The power values are shown in descending order, therefore, the first antenna (or beam) is the most powerful. The index means beam index or antenna index, depending on whether the signal was transformed to the beamspace domain or not. For the antenna signal transformation to the beamspace domain, we employed two standard approaches: discrete Fourier transform (DFT) and singular value decomposition (SVD). The last one is more rarely employed because of a higher complexity of signal transformation to the beamspace and a higher sensitivity to the channel aging effect \cite{bychkov2021data, access}.

\begin{table}[t!]
    \renewcommand{\arraystretch}{1.1}    
        \begin{tabular}{|p{2.3cm}|p{1.0cm}||p{2.0cm}|p{1.8cm}|}\hline
            \textbf{Parameter}    & \textbf{Value}  &  \textbf{Parameter}& \textbf{Value}   \\ \hline
            \textbf{Carrier frequency}   &  3.5GHz  &  \textbf{Modulation} & QAM256 \\ 
            \textbf{Subcarrier spacing}  &  30kHz   &  \textbf{N of scenarios} & 140 \\ 
            \textbf{BS height}  & 25m               &  \textbf{N of noise seeds} & 16 \\ 
            \textbf{UE height} &  1.5m              &  \textbf{RB number} & 16 \\ 
            \textbf{Vertical antennas spacing}      & 0.9$\lambda$   &  \textbf{SNR} & 20dB  \\ 
            \textbf{Horizontal antennas spacing}    & 0.5$\lambda$ &   \textbf{Channel model}  & 3GPP-3D, Berlin, Dresden \\
            
            \textbf{N of BS antennas}   &   64  & \textbf{Channel type}   &  NLOS  \\ 
            \textbf{N of UE antennas}  &   2 & \textbf{User speed} & 5 km/h \\ 
            \textbf{UE number}  &  1 or 4 & \textbf{Beamspace type}   &  DFT and SVD  \\ 
            \textbf{N of subcarriers per RB ($N_{12}$)}   &   12  & \textbf{Beamspace size $\bm{N_{beam}}$} & 16 or 32 \\ 
            \hline
        \end{tabular}
        \caption{Simulation parameters.}
        \label{tab:params}
\end{table}

As we can see, the received power is almost independent of the antenna index, because antennas are omnidirectional. In this case, the equation \eqref{1user} is valid. Unfortunately, the signal dimensionality can't be reduced without performance loss, as almost all antennas contain user signals. 

In contrast to the antenna domain, in the beamspace domain, the signal becomes sparse. In the DFT case, the first 16 beams contain 93$\%$ of power, while for the SVD it contains 99.7$\%$ of total power. Since the user power strongly depends on the beam index, mantissa length ($\delta_Y$ in equation \eqref{1user}) should also be beam-dependent.

\subsection{Surrogate optimization}
To find the optimal mantissa lengths for each beam, we employed a surrogate optimization because it is fast, and can work with black-box models \cite{wang2014general}. 

Surrogate modeling is a technique used in optimization to approximate the objective function when direct evaluation is computationally expensive. The algorithm alternates between constructing a surrogate model using sampled points and searching for the minimum of the objective function. The surrogate is typically created using radial basis functions. During the search phase, the algorithm evaluates a merit function based on the surrogate and distances between points. This approach efficiently balances exploration and speed, updating the surrogate iteratively to refine the search for optimal solutions. The algorithm has been proven to converge to a global solution for objective functions on bounded domains.

For surrogate optimization, we employed the loss function consisting of the following three components: 
\begin{equation}\label{loss}
L = L_{1} + \alpha \cdot L_{2} + \beta \cdot L_{3},
\end{equation}

The first term is responsible for the FH compression and is calculated as an overall length of $N_{beam}$ mantissas:
\[
L_{1} = \frac{1}{N_{beam}}\sum_{i=1}^{N_{beam}} B_i , 
\]
where $B_i$ is the bitwidth of non-uniform quantizer at $i$-th beam (simply mantissa length at $i$-th beam). 

The second term is required to guarantee the actual $EV\!M$ to be close to the desired $EV\!M_{0}$ value as follows:
\[
 L_{2} = \max \left( \frac{EV\!M}{EV\!M_{0}}-1,0\right),
\]

The last term helps to avoid overfitting as the mantissa length should not increase with the beam index according to the beamspace selection strategy:
\[
L_{3} = \sum_{i=1}^{N_{beam}-1} \max \left(B_{i+1}-B_{i},0 \right).
\]
Let us remind, that beams are sorted in advance according to the received signal power.

Finally, hyperparameters $\alpha \approx 2^5$ and $\beta \approx 2^{-1}$ represent scaling coefficients to adjust the ranges. They balance the trade-off between losses $L_{1..3}$ . 

\subsection{Compression ratio}

In practice, antenna transformation to the beamspace domain provides significant signal compression because of reduced dimensionality and flexible mantissa length. The compression ratio is calculated as a ratio of the overall number of bits in the fixed-point format in the antenna domain and the overall number of bits in the common-exp package in \Fig{common_exp} in the beamspace domain. Thus, it can be calculated as follows:  
\begin{equation}\label{compression_ratio}
CR = \frac{2N_{SC}B_{FP}N_{RX}}{B_{EXP}N_{beam}N_{RB}+2N_{SC}\sum_{i=1}^{N_{beam}} B_i},
\end{equation}
where $B_{EXP}$ is the exponent length, $B_{i}$ is the mantissa length for the $i$-th beam, and $B_{FP}$ is the symbol bitwidth in the fixed-point format (before compression).

\section{Simulation results}
\label{sec:experiments}

For simulations, we employed the Quadriga 2.0 software \cite{Quadriga} with parameters defined in Table \ref{tab:params} to generate a realistic non-line-of-sight (NLOS) channel. The propagation channel matrix was assumed to be ideally estimated. For the MIMO detection we employed the MMSE algorithm \eqref{MMSE} implemented in the single precision format in the beamspace domain. The Cholesky decomposition \cite{OurCholBound} was utilized for matrix inversion. 

We utilized the nonlinear Gaussian quantizer to compress mantissa, as the distribution is close to the Gaussian one \cite{Efficient}. For tests, the common exponent value of $B_{EXP}=4$ bits was used, as shown in \Fig{common_exp}. For fixed mantissa length, we utilized $B=6$ bits and calculated the corresponding EVM value (compression with a fixed mantissa length). Then the surrogate optimizer utilized the range $1...10$ bits to find the mantissa length values ($B_i$) to minimize the loss function \eqref{loss}. As a result, the overall FH bitwidth is minimized for the same EVM (compression with a flexible mantissa length).

\subsection{Fixed mantissa length - error prediction}

\begin{figure}[t!]
\centering
\includegraphics[width=1.00\columnwidth]{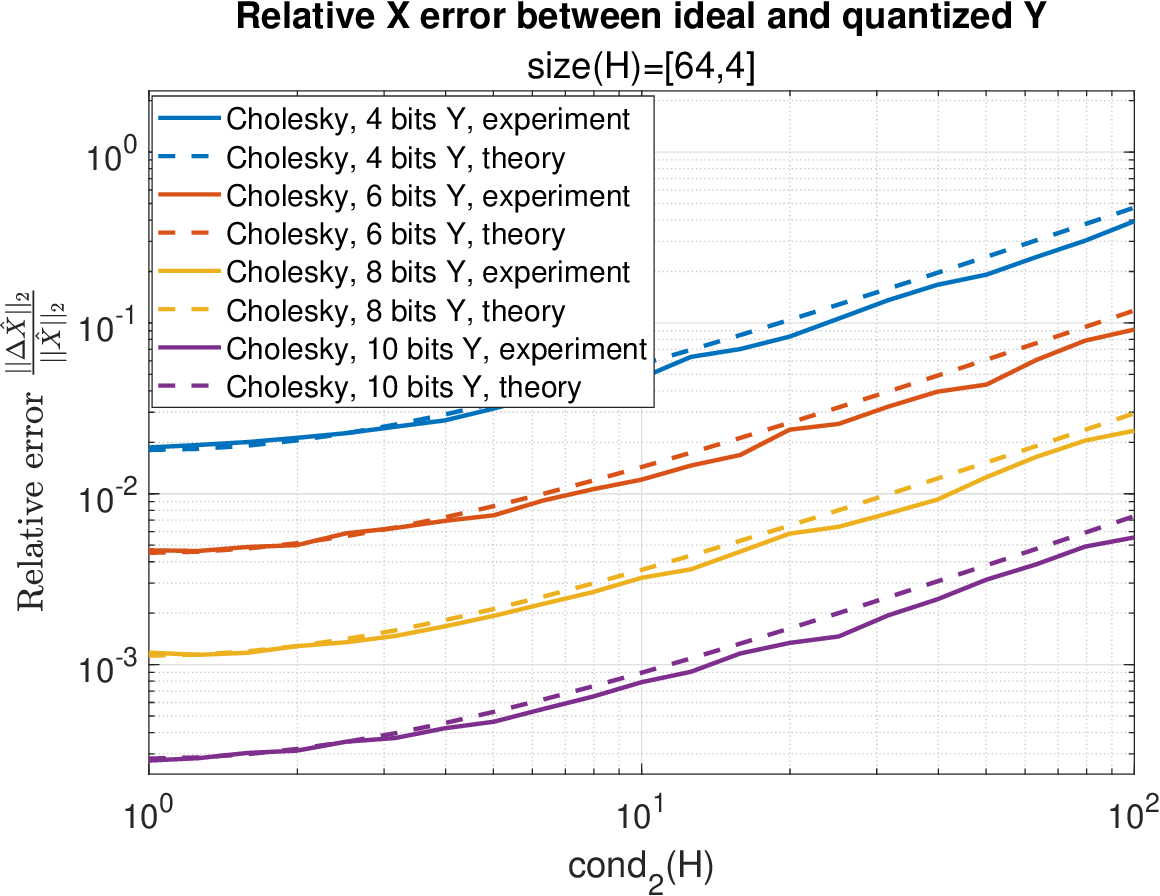}
\caption{Roundoff error after MIMO detector depending on the condition number (antenna domain).
}
\label{Error_64x4}
\end{figure}

The MIMO detection error caused by the FH compression is presented in \Fig{Error_64x4} depending on the condition number of the propagation channel matrix for a scenario with 4 users. Curves correspond to FH mantissa lengths 4, 6, 8, and 10 bits. One can see the error predicted by eq. \eqref{mu_common_exp} closely matches the simulated one.

\subsection{Flexible mantissa length - better compression}

For simulations, we employed a single-user scenario. The constant mantissa length of $B=6$ (baseline) results in the EVM=$1.6\%$ for the SNR value of $20$dB per antenna. Then we trained mantissa lengths depending on the beam index according to eq. \eqref{loss}. 

\begin{figure}[t!]
\centering
\includegraphics[width=1.00\columnwidth]{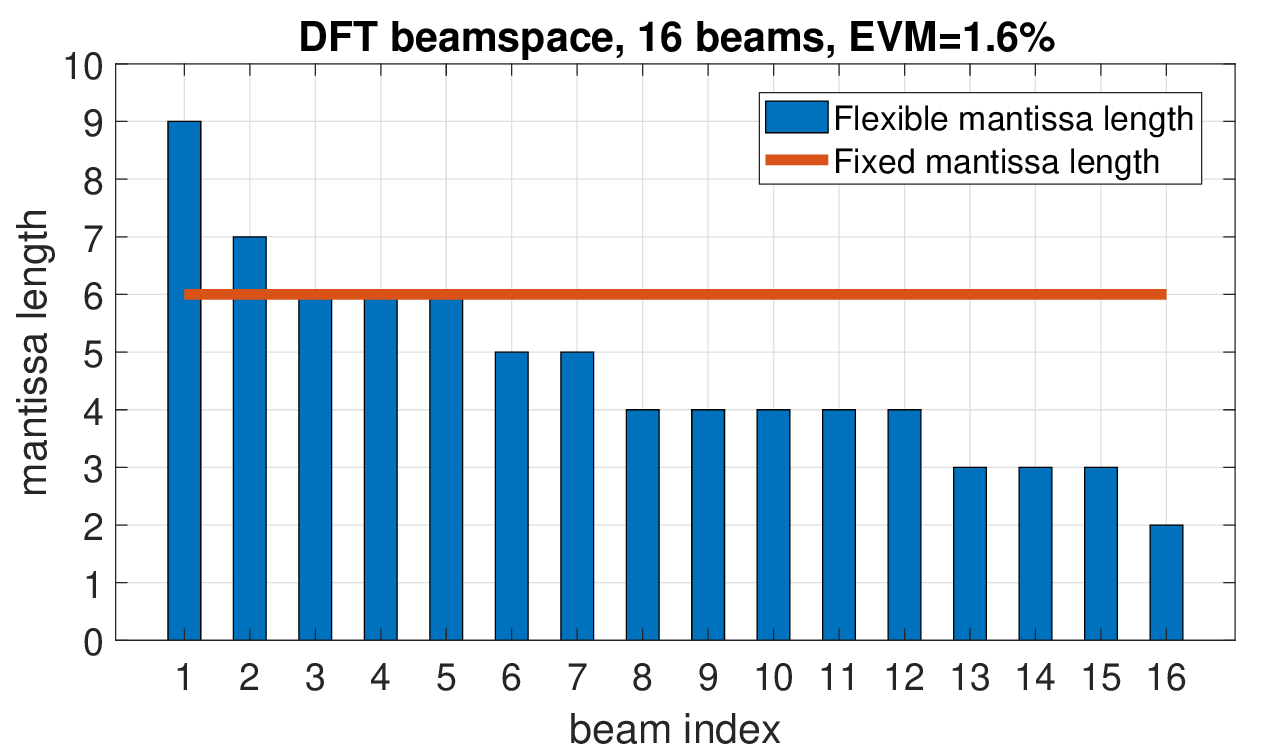}
\caption{Mantissa length distribution for the DFT beamspace (single user).
}
\label{power_distribution1}
\end{figure}

\begin{figure}[t!]
\centering
\includegraphics[width=1.00\columnwidth]{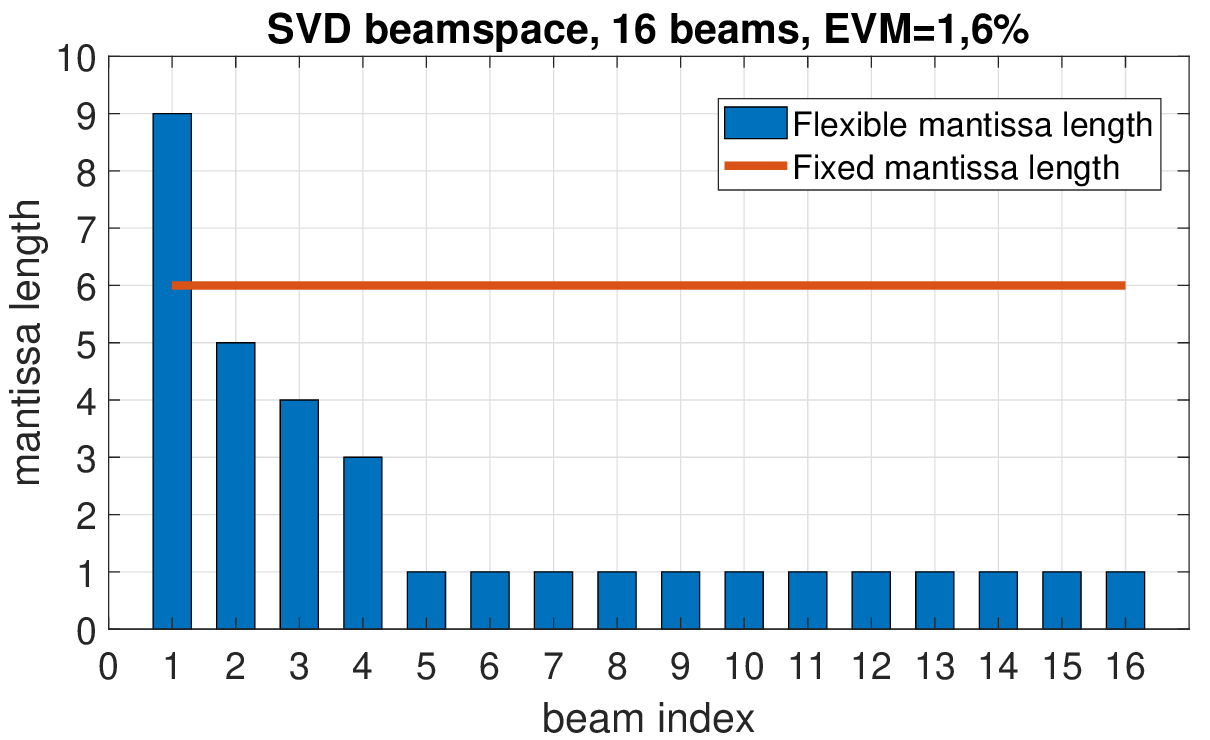}
\caption{Mantissa length distribution for the SVD beamspace (single user).
}
\label{power_distribution2}
\end{figure}

The optimized mantissa lengths are presented in \Fig{power_distribution1} for the DFT-based beamspace and in \Fig{power_distribution2} for the SVD-based beamspace. Let us remind that all beams are sorted according to the descending received signal power. 

Thus, despite the additive noise power is the same in each beam, the most powerful beams require a higher bitwidth because of a higher SNR there that can be easily derived for a single user case from eq. \eqref{MMSE}. The user power in the least powerful beams is almost negligible, therefore, the corresponding values of the weight vector $W$ are small, and the quantization noise power there is not critical. Thus, the bitwidth in beams with high indexes was reduced without performance degradation. 

The compression ratio was calculated using equation \eqref{compression_ratio}. Its values are presented in Table \ref{CR} for online and offline training modes. Online mode means the mantissa lengths should be optimized for each particular scenario ``on the fly'' (upper compression bound). While the offline mode means the mantissa lengths have been optimized in advance and can be applied to any received signal (practical approach). The corresponding mean mantissa lengths are shown in Table \ref{Mantissa}.

\begin{table}[t!]
\centering
    \renewcommand{\arraystretch}{1.2}    
        \begin{tabular}{|c|c|c|c|c|}\hline
            \multirow{2}{*}{MODE} & \multicolumn{2}{c|}{16 beams} & \multicolumn{2}{c|}{32 beams}  \\ 
             &  DFT  &  SVD  & DFT  &  SVD \\  
            \hline
            \textbf{Online trained}    & 16.5  & 31.0  & 8.7  &  19.2 \\ 
            \hline
            \textbf{Offline trained}    & 12.9 &   28.2  & 7.7  &  18.1\\
            \hline
            \textbf{Fixed length} & \multicolumn{2}{c|}{10.4} & \multicolumn{2}{c|}{5.2}\\ 
            \hline
        \end{tabular}
        \caption{Compression ratio (single user)}
        \label{CR}
\end{table}

\begin{table}[t!]
\centering
    \renewcommand{\arraystretch}{1.2}    
        \begin{tabular}{|c|c|c|c|c|}\hline
            \multirow{2}{*}{MODE} & \multicolumn{2}{c|}{16 beams} & \multicolumn{2}{c|}{32 beams}  \\ 
             &  DFT  &  SVD  & DFT  &  SVD \\ 
            \hline
            \textbf{Online trained}    & 3.7  &  1.9 & 3.5  &  1.5 \\ 
            \hline
            \textbf{Offline trained}    & 4.8 &   2.1  & 4.0  &  1.6\\
            \hline
            \textbf{Fixed length} & \multicolumn{4}{c|}{6} \\ 
            \hline
        \end{tabular}
        \caption{Mean mantissa bitwidth (single user)}
        \label{Mantissa}
\end{table}

\section{Conclusion}


The proposed algorithm can accurately predict the MMSE detector error caused by the fronthaul compression. Thus, depending on the number of users and beams as well as on the condition number of the propagation channel matrix, the fronthaul bitwidth can be lowered without performance loss. This allows for a higher compression ratio without compromising the required EVM for each particular scenario.

The proposed flexible mantissa length (depending on the beam index) can be pre-trained to improve the fronthaul compression by leveraging the propagation channel sparsity. Simulations with the DFT-based beamspace show the compression ratio is increased by $20...33\%$ compared with a fixed mantissa length while maintaining the same detection error.




\label{sec:conclusion}



\bibliographystyle{IEEEtran}
\bibliography{main.bib}

\end{document}